\documentstyle[epsfig]{mn}
\input{psfig}
\begin{document}
\def\ltsima{$\; \buildrel < \over \sim \;$}
\def\simlt{\lower.5ex\hbox{\ltsima}}
\def\gtsima{$\; \buildrel > \over \sim \;$}
\def\simgt{\lower.5ex\hbox{\gtsima}}

\title[Soft X-ray spectroscopy of Compton-thick Seyfert~2 galaxies with 
BeppoSAX]
{Soft X-ray spectroscopy of Compton-thick Seyfert~2 galaxies with BeppoSAX}

\author[M. Guainazzi, et al.]
{M. Guainazzi$^1$, G.Matt$^2$, L.A.Antonelli$^3$,
L.Bassani$^4$, A.C.Fabian$^5$, R.Maiolino$^6$ \and A.Marconi$^6$,
F.Fiore$^{3,7}$, K.Iwasawa$^5$, L.Piro$^8$ \\ ~ \\
$^1$Astrophysics Division, Space Science Department of
ESA, ESTEC, Postbus 299, NL-2200 AG Noordwijk, The Netherlands \\
$^2$Dipartimento di Fisica ``E.Amaldi'', Universit\'a degli Studi ``Roma 
Tre'', Via della Vasca Navale 84, I-00146 Roma, Italy \\
$^3$Osservatorio Astronomico di Roma, Via dell'Osservatorio, I-00144 
Monteporzio Catone, Italy \\
$^4$Istituto Tecnologie e Studio delle Radiazioni Extraterrestri, C.N.R., Via 
Gobetti 101, I-40129 Bologna, Italy \\
$^5$Institute of Astronomy, University of Cambridge, Madingley Road, Cambridge 
CB3 0HA, United Kingdom \\
$^6$Osservatorio Astrofisico di Arcetri, Largo E.Fermi 5, I-50125 Firenze, 
Italy \\
$^7$BeppoSAX Science Data Center, Via Corcolle 19, I-00131 Roma, Italy \\
$^8$Istituto di Astrofisica Spaziale, CNR, Via Fosso del Cavaliere, I-00133 
Roma, Italy
}

\maketitle
\begin{abstract}

We present a X-ray spectroscopic study of the bright Compton-thick
Seyfert~2s NGC~1068 and Circinus Galaxy, performed with BeppoSAX.
Matt et al. (1997 and 1998)
interpreted the spectrum above 4~keV as the
superposition of Compton reflection and warm plasma scattering
of the nuclear radiation.
When this continuum is extrapolated downwards to 0.1~keV, further
components arise. The NGC~1068 spectrum
is rich in emission lines, mainly due to K$_{\alpha}$ transitions of
He-like elements from oxygen to iron, plus a K$_{\alpha}$ fluorescent line
from neutral iron. If the ionized lines originate in the warm scatterer,
its thermal and ionization structure must be complex. From the
continuum and line properties, we estimate a column density,
${\rm N_{warm}}$, of the warm scatterer less than
 a few$\times10^{21}$~cm$^{-2}$. In Circinus Galaxy, the absence
of highly ionized iron is consistent with a scattering medium with
${\rm U_X \simlt 5}$ and ${\rm N_{warm} \sim}$ a few$\times10^{22}$~cm$^{-2}$.
In both cases the neutral iron line is most naturally explained as
fluorescence in the medium responsible for the Compton reflection continuum.
In NGC~1068 an optically thin
plasma emission with ${\rm kT \simeq}$500~eV and strongly sub-solar
metallicity is required, while such a component is only marginal in Circinus
Galaxy. We tentatively identify this component as
emission of diffuse hot gas in the nuclear starbursts. Possible causes
for the metal depletion are discussed.
\end{abstract}


\section{Introduction}

In the last fifteen years, a wide consensus 
has gathered around the idea that the nucleus
and the Broad Line Region (BLR) in Seyfert~2s are hidden to us 
by intervening matter along the line of sight (see Antonucci 1993 for a review
on unification models). While column densities 
${\rm \approx 10^{22}}$--${\rm 10^{24}}$~cm$^{-2}$ were
measured by Ginga in most Seyfert~2s
(Awaki et al. 1991; Smith \& Done 1996), so confirming this hypothesis,
no significant absorption in excess 
of the Galactic contribution was detected in some objects. 
A key to explain the nature of these apparently odd sources
came from the evidence that they exhibit intense iron lines,
strongly suggesting that the observed X--ray emission is
due to scattering 
of the nuclear radiation, which in turn is completely obscured. In fact, 
if the column density of the absorbing matter exceeds
$10^{24}$~cm$^{-2}$), the nucleus is invisible 
up to at least 10~keV (and then no absorption would be directly observed).
At least two possible candidates exist as reflectors: the inner side
of the molecular torus (``cold reflector'')
and the hot plasma responsible for the scattering
of optical broad lines (``warm mirror'' or ``warm reflector'').

In the former case, the X-ray spectrum should be dominated by a
``bare'' Compton-reflection
component (George \& Fabian 1991; Ghisellini et al. 1994; Krolik et al. 1994)
and an intense (equivalent width ${\rm EW \simgt
1}$~keV) K$_{\alpha}$ fluorescent line from neutral or mildly
ionized iron (Matt et al. 1996a, MBF96 hereinafter).
This is indeed what has been observed
in a handful of objects so far: NGC~1068 (the archetypical
Seyfert~2 galaxy: Ueno et al. 1994; Iwasawa et al. 1997;
Matt et al. 1997a, M97 hereinafter),
NGC~6240 (Iwasawa \& Comastri 1998); Circinus Galaxy (Matt et al. 1996b,
Matt et al. 1999; M99 hereinafter); NGC~7674 (Malaguti et al.
1998); Mkn~3 (Turner et al. 1997a; Cappi et al. 1999). These objects constitute
the set of so--called ``Compton-thick'' sources.
Recent results from a BeppoSAX study of an optically-selected sample
suggest that they are far more common than previously thought 
due to selection biases in the Seyfert~2 samples observed in X-rays
before (Maiolino et al. 1998).

The X-ray continuum scattered by the warm material 
is instead a fainter replica of the nuclear
non-thermal continuum (if self-absorption effects are not important),
plus a set
of emission lines from highly ionized heavy elements, which are 
produced by 
fluorescence/recombination and/or resonant scattering in the ionized matter
(MBF96; Netzer 1996;
Netzer et al. 1998, NTG98 hereinafter).

The X-ray observations, performed so far, suggest that both mechanisms 
are usually at work. Soft excesses above the extrapolation of the
intermediate X-ray 2--10~keV power-laws have been revealed in
21 out of 25 Seyfert~2s observed by ASCA (Turner et al. 1997b), and in all 
``Compton-thick'' sources 
(Turner et al. 1997b; Iwasawa \& Comastri 1998; Matt et al. 1996b).
Emission lines from He-, H-like iron and lighter elements
have been detected in 4 out of 7 Compton-thick sources (Matt et al. 1996b;
Turner et al. 1997b).
Scattering models can generally account well for these soft excesses.
However, a description
of these features in terms of optically thin thermal emission often yields
statistically comparable results. This is not surprising.
Some studies indicate that the host
galaxies of Seyfert~2s have high levels of star formation 
(Maiolino \& Rieke 1995) and several
examples of nuclear starbursts occurring in Seyfert~2s
are known. The average contribution of starbursts to the
0.5-4.5~keV flux was estimated to be 60\% in the sample of Seyfert~2s observed
by ASCA (Turner et al. 1997b), although admittedly ASCA could not
disentangle unambiguously
the starburst from the scattered emission. In only
one case a significant contribution from starburst could be ruled out
(Mkn~3; Turner et al. 1997a).

High spatial resolution is required to separate the contribution 
of different spectral components and to 
perform spatially-resolved spectroscopy. Hopefully, this capability will be 
provided in the near future  by the detectors on board {\it Chandra} and XMM.
In the meantime, a wide enough spectral coverage 
can be exploited to separate 
the cold and warm reflection components 
above 3~keV and to quantify any contribution
to the soft X-ray emission from a starburst component. 
The Italian-Dutch satellite BeppoSAX (Boella et al. 1997a),
whose scientific payload covers the energy range between 0.1 and
200~keV, is the presently operating mission best suited for this purpose. 
For example it has allowed for the first time the measurement
of the relative contributions
of the cold and warm mirrors in NGC~1068,
confirming the complex nature of the reflector, first
suggested by ASCA spectroscopy of the iron line
complex (Iwasawa et al. 1997). In this {\it paper} we will focus
on the soft X-ray properties of NGC~1068 and Circinus Galaxy, which
were observed by BeppoSAX during
a program of spectral survey of Compton-thick Seyfert~2s.

Both sources are very pertinent to the above discussion.
In both of them, broad optical lines have been detected
in polarized light (the degree of polarization being
$\simeq$16\% in NGC~1068, Antonucci
\& Miller 1985;  and $\simeq$2\% in Circinus Galaxy, Oliva et al. 1998).
Both galaxies are site of strong starburst activity. In NGC~1068
the bulk of this activity is concentrated in a ring of 
approximately 1~kpc size [{\it i.e.}: 15-16'' at 
the distance of 14.4~Mpc (Tully 1988)],
which protrudes a bar towards the nucleus
(Scoville 1988). The HRI image revealed that half of the X-ray emission
at 0.8~keV comes from an extended region of
$\simeq 13$~kpc scale (Wilson et al. 1992). It is worth remembering
that NGC~1068 hosts a ``water maser'' source (Claussen et al. 1984),
probably coincident with the inner region of a dusty, nearly edge-on
torus at a distance of $\simeq$0.4~pc from the nucleus (Greenhill et al.
1996). 

Circinus Galaxy also exhibits a strong and variable ${\rm H_2}$
maser emission (Greenhill et al. 1997),
which instead may be produced in a
$\simlt$10~AU Keplerian disk around the nucleus. Enhanced star formation
 activity in the shape of a ring is present
on a $\simeq$200~pc scale (corresponding to
$\simeq$10'' at the distance of 4~Mpc; Marconi et al. 1994).
Although the total luminosity of
the starburst within a few hundred pc is comparable to the intrinsic
luminosity of the Seyfert nucleus, only 2\% of it is radiated
within the inner 12~pc (Maiolino et al. 1998a).

\section{Observations, data reduction and preparation}

The Italian-Dutch satellite BeppoSAX (Boella et al. 1997a) carries
four co-aligned Narrow Field Instruments. Two of them are gas scintillation
proportional counters with imaging capabilities: the Low Energy
Concentrator Spectrometer (LECS, 0.1--10~keV, Parmar et al. 1997)
and the Medium Energy Concentrator Spectrometer (MECS, 1.8--10.5~keV,
Boella et al. 1997b). They have an energy resolution of $\simeq$8\%
at 6~keV (MECS) and of $\simeq$4\% at 1~keV (LECS).
The other two instruments are collimated 
detectors, mounted on a rocking system to allow
a continuous monitoring of the background: the High Pressure Gas
Scintillator Proportional Counter (HPGSPC, 4-120~keV, Manzo et al. 1997)
and the Phoswitch Detector System (PDS, 13-200~keV). The HPGSPC is
tuned for spectroscopy of bright sources with good energy resolution,
while the PDS possesses an unprecedented sensitivity in its energy
bandpass. Only LECS, MECS and PDS data will be considered in this {\it paper},
since the HPGSPC failed to detect both sources.

The NGC~1068 observation was interrupted after $\simeq$70\% of the scheduled
exposure time and completed about one year later.
The log of the observations is reported in Table~\ref{tab5}.
\begin{table*}
\centering
\begin{footnotesize}
\caption{BeppoSAX observations log. ${\rm T_{exp}}$ and ${\rm CR}$
are the total effective exposure time and count rate in the 0.1--4~keV,
1.8--10.5~keV and 13--200~keV bands for the LECS, MECS and PDS, respectively.}
\label{tab5}
\vspace{0.05cm}
\begin{center}
\begin{tabular}{lllcccccc} \hline \hline
Source & Start time (UTC) & End time (UTC) & ${\rm T^{LECS}_{exp}}$ & ${\rm 
CR^{LECS}}$ & ${\rm T^{MECS}_{exp}}$ & ${\rm CR^{MECS}}$ & ${\rm 
T^{PDS}_{exp}}$ & ${\rm CR^{PDS}}$ \\ 
& & & (s) & (s$^{-1}$) & (s) & (s$^{-1}$) & (s) & (s$^{-1}$) \\ \hline
NGC~1068 (1) & 30/12/96 08:47:25  & 03/01/97 05:27:50 & 61497 & $0.1163 \pm 
0.0014$ & 100150 & $0.0872 \pm 0.0011$ & 62493 & $0.18 \pm 0.02$ \\
NGC~1068 (2) & 11/01/98 09:52:36 & 12/01/98 08:07:50 & 15408 & $0.110 \pm 
0.03$ & 37331 & $0.0664 \pm 0.0015$ & 17657 & $0.15 \pm 0.06$ \\ 
CG & 13/3/98 06:33:01 & 17/3/98 06:21:05 & 83853 & $0.0162 \pm 0.0005$  & 
137700 & $0.0962 \pm 0.0009$ & 63200 & $2.01 \pm 0.04$ \\ \hline \hline
\end{tabular}
\end{center}
\end{footnotesize}
\end{table*}
Data were reduced according to the same prescriptions as in M97.
The only difference regards the PDS data, on which a crystal
temperature dependent Rise Time threshold was applied
during the screening procedure. 
By reducing the instrumental background
by $\sim 50\%$, this method allowed an improvement in the S/N ratio from 4.5
to 8$\sigma$. We assume hereinafter
that this also implies a decrease by a factor
0.92 of the effective area in comparison to the publicly available matrix
(September 1997 release, Fiore et al. 1998), which were used throughout.
Background spectra for the imaging
instruments were extracted from blank field event files, using the
same region in detector coordinates where the sources lie. PDS background
subtracted spectra were obtained by plain subtraction of the
``off-source'' from the ``on-source'' ones.
In all  spectral fits, multiplicative factors
were included
to account for the absolute flux cross calibration misalignment
between the
BeppoSAX detectors (Grandi et al. 1997; Haardt et al. 1998).
The LECS vs. MECS
factor was left free to vary as a free parameter in the fits,
and turned out to be $\simeq$0.9.
The PDS vs. MECS factor was fixed to 0.75.
The results are not substantially affected
by the residual $\simlt 5\%$ uncertainty on the last parameter
(Fiore et al. 1998).

In the following, errors are at the 90\% confidence level
for one interesting parameter ({\it i.e.}: ${\rm \Delta \chi^2 = 2.71}$);
energies are quoted in the source rest frame; 
${\rm H_0 = 50}$~km~s$^{-1}$~Mpc$^{-1}$, and cosmic abundance after
Anders \& Grevesse (1989) are assumed.

\section{NGC~1068}

Source spectra for NGC~1068
were extracted from circular regions of 8' and 6' 
radius for the LECS and MECS, respectively.
The spectral analysis was performed in two steps.
First, we re-analyzed the high--energy ({\it i.e.} above
4~keV) spectrum, to check the results of M97 after
the improved
PDS data screening algorithm and also to include the 1998 data set.
Once the high--energy continuum shape was estimated, we tackled
the task of describing the broadband spectrum.
We have not found any difference in the spectral parameters larger than the
statistical uncertainties between the 1996 and 1998 observations in any
of the spectral model employed in this {\it paper}. The spectral
analysis have therefore been performed on their
union\footnote{The analysis of the total spectrum has been performed
on the LECS and PDS summed spectra, while the MECS spectra of the
two observations have been maintained separated and fitted simultaneously
given the different number of operative MECS units (three versus two) between
the two observations.} (``total'' dataset  hereinafter).

\subsection{${\rm E > 4}$~keV}

In the analysis of the NGC~1068 spectra, we have followed the same
approach as in M97.
The MECS and PDS spectra above 4~keV have been fitted simultaneously with a
model composed by a Compton reflection continuum (model {\verb!pexrav!}
in {\sc Xspec}, Magdziarz \& Zdziarski 1995)
and a power-law (describing the warm reflector component), 
both of them absorbed by the Galactic column density
(${\rm N_H = 3.1 \times 10^{20}}$~cm$^{-2}$, Dickey \& Lockman 1990).
The photon index of the
power--law and of the illuminating primary continuum,
which gives rise to the Compton
reflected component, were tied together. No cut-off has been
assumed in the primary component. To describe the iron line
complex we first tried a system of
three narrow ({\it i.e.}: intrinsic width
$\sigma$ equal to 0) Gaussian lines; the centroid energy ${\rm E}$
of one of these components
was held fixed to 6.4~keV ({\it i.e.}: fluorescence from neutral or
mildly ionized iron) while  the other two were left free to vary.
The addition of a further narrow line
yields a $\Delta \chi^2 = 8.3$ for
two degrees of freedom, significant at more than 97.5\% level of confidence,
giving a line energy of $\simeq 8.1$~keV.
The quality of the fit is very good (${\rm \chi^2 = 165.7/179}$~dof) and
the photon spectral index
($\Gamma \simeq 2.0 \pm 0.2$)
is consistent with that typically observed in Seyfert 1
galaxies (Nandra \& Pounds 1994; Nandra et al. 1997).
The best--fit solution requires the centroid energy of the K$_{\alpha}$ ionized
iron lines to be $\simeq 6.56^{+0.12}_{-0.07}$~keV
(corresponding
to Fe{\sc xx}$^{+\hbox{\sc iii}}_{-\hbox{\sc ii}}$) and
$6.96^{+0.13}_{-0.08}$~keV ({\sc Fe xxvi}),
with Equivalent Widths (EW) against the total continuum of
740 and 820~eV, respectively. The
8.1~keV line has a centroid energy consistent with the
K$_{\beta}$ transition from H-like iron (${\rm E_c = 8.1 \pm 0.2}$~keV)
and its ratio with the 6.96~keV line
is $0.18 \pm 0.12$. 
The MECS does
not have enough energy resolution to
allow a totally unambiguous deconvolution of the iron line complex
in this source.
Comparably good fits (albeit formally slightly worse)
can be obtained also if one assumes
a system of Fe{\sc i}+6.61~keV+6.86~keV lines (as suggested
by Iwasawa et al. 1997; ${\rm \chi^2 = 173.6/180}$~dof)
or of Fe{\sc i}+Fe{\sc xxv}+Fe{\sc xxvi}
(${\rm \chi^2 = 171.5/180}$~dof). The
8.1~keV line is required at comparable statistical
significance in all the above deconvolutions. 
The relative weight of the K$_{\alpha}$ iron components varies
significantly in the three models. In the following we will
assume the last one as our baseline as it is the most physically reasonable. 
The properties
of the lines in this scenario are reported in Table~\ref{tab3}.

\begin{figure}
\begin{center}
\epsfig{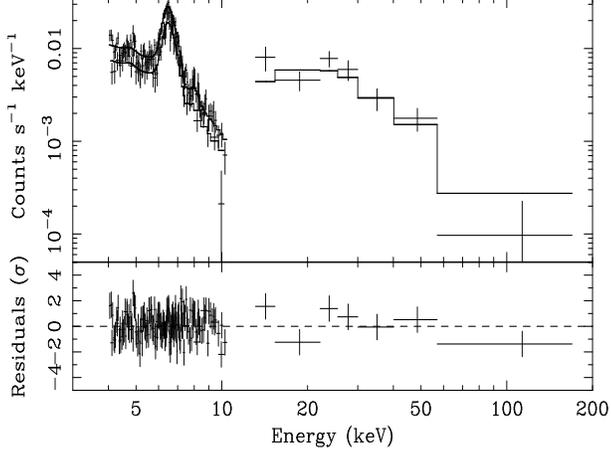}
\caption{NGC~1068 spectra ({\it upper panel}) and residuals
in units of standard deviations ({\it lower panel}) when
the best-fit double-reflector
model is applied to the MECS/PDS spectra above 4~keV}
\label{fig2}
\end{center}
\end{figure}

In Figure~\ref{fig2} the best--fit model and residuals
photon spectrum is shown for the total dataset. It will be referred to
as the {\bf ``double-reflector''} model hereinafter, whereas the {\bf double-reflector
continuum} will refer to the above model without the iron emission line
complex.

\subsection{Broadband spectrum}

\begin{figure}
\begin{center}
\epsfig{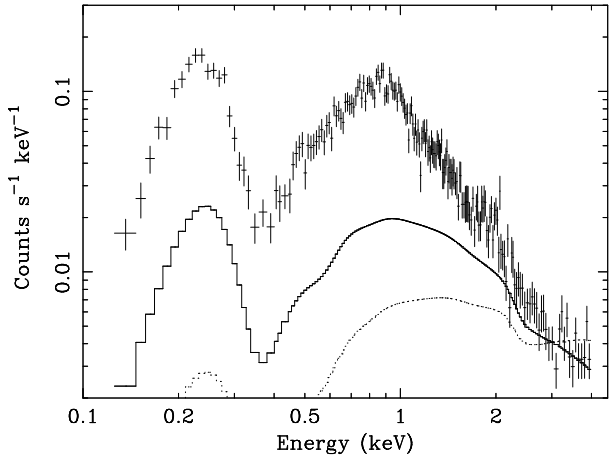}
\caption{Extrapolation of the high--energy best--fit double-reflector model
({\it solid line}) in the LECS energy bandpass. Note the
huge soft excess observed ({\it crosses}). The
extrapolation of a power--law having the same flux at 4~keV but spectral 
index $\Gamma = 1.2$ (as observed by Ueno et al. (1994, {\it dotted line})
is also shown for comparison}
\label{fig3}
\end{center}
\end{figure}

In Figure~\ref{fig3}, the extrapolation of the double-reflector model in
the LECS band (below 4~keV) clearly shows the presence of a huge soft excess.
Although this component has already been pointed out by 
several authors in the past
(Marshall et al. 1993; Ueno et al. 1994),
BeppoSAX
unprecedented broadband coverage allows for the first time a
self-consistent simultaneous study of {\it both} the hard emission
and the soft excess properties.
It is worth stressing that the soft excess is present despite the
fact that the extrapolated spectrum is now much steeper than
assumed so far. Previous studies have in fact calculated a mean
intermediate X-ray spectrum and estimated the soft excess above it
(cf. Figure~\ref{fig3}). Now we can deconvolve the  {\it observed}
flat spectrum in
two physical components. 

\begin{figure}
\begin{center}
\epsfig{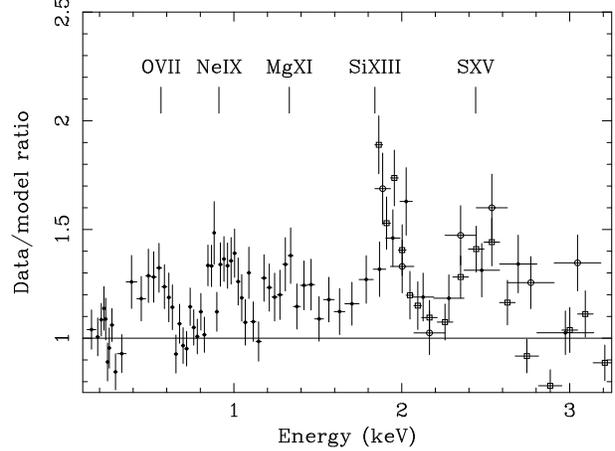}
\caption{Data/model ratio when the double reflector model plus a
thermal optically thin
plasma is applied to the
LECS ({\it filled circles}), MECS ({\it empty squares}, 1996
dataset; {\it empty circles}, 1998 dataset) and PDS spectra
(not shown) simultaneously. The labels indicate the likely
identification of the observed lines (marked at the expected energies
after taking into account the redshift of the source).
Each data point has a signal-to-noise ratio $> 10$}
\label{fig4}
\end{center}
\end{figure}

We have modeled the soft excess by adding  to the double-reflector model
a thermal plasma emission
(model {\verb!mekal!} in {\sc Xspec}), or
a bremsstrahlung or a  power-law.
Although the quality of the fit improves dramatically in each case,
local wiggles can still be seen both in the LECS and in the MECS spectrum
(see Figure~\ref{fig4}),
suggesting that localized emission features are required by the data
as well.
We have therefore followed the approach of adding
narrow Gaussian lines, until statistically required by the fit
at the 99\% significance level. This yields a system of five
soft X-ray lines in the range 0.5--2.5~keV. The best-fit continuum and
line parameters are shown in Table~\ref{tab1} and~\ref{tab3}, respectively.
\begin{table}
\begin{footnotesize}
\centering
\caption{Best-fit results obtained for NGC1068 when an extra component
(indicated in the first row) is added to the double-reflector model
in order to describe the observed soft excess.
{\it po}~=~power-law; {\it br}~=~thermal
bremsstrahlung; {\it mk}~=~optically thin plasma}
\label{tab1}
\vspace{0.05in}
\begin{center}
\begin{tabular}{lccc} \hline \hline
Parameter & {\verb!po!} & {\verb!br!} & {\verb!mk!} \\ \hline
${\rm N_H}$ ($10^{20}$~cm$^{-2}$) & $5.0 \pm 0.4$ & $3.1 \pm 0.3$ & $2.4 \pm 
0.4$ \\
${\rm \Gamma_{hard}}$ & $2.30 \pm^{0.11}_{0.09}$ & $2.12 \pm^{0.17}_{0.16}$ & 
$2.13 \pm 0.17$ \\
${\rm \Gamma_{soft}}$ & $3.21 \pm^{0.13}_{0.10}$ & &  \\
${\rm N_{sc}/N_{pexrav}}$ (\%) & $< 1.2$ & $7 \pm^3_4$ & $ 7 \pm 4$ \\
${\rm kT}$ (eV) & & $410 \pm^{40}_{30}$ & $440 \pm 50$ \\
${\rm A_Z}$ (\%) & & & $2.8 \pm 1.6$ \\
$\chi^2/$dof & 513.0/442  & 453.7/440 & 451.3/441 \\ \hline \hline
\end{tabular}
\end{center}
\end{footnotesize}
\end{table}
\begin{table}
\begin{footnotesize}
\centering
\caption{X-ray lines detected in NGC~1068. Centroid energies are
in keV, EW in eV. ${\rm EW_{tot}^{(i)}}$ is the equivalent
width against the total continuum, ${\rm E_{sc}^{(i)}}$ against
the warm scattered continuum only, ${\rm E_{cr}^{(i)}}$ against the
Compton-reflection continuum only. The third column indicates the most likely
identification. In brackets the line fluxes are quoted in units of
photons~cm$^{-2}$~s$^{-1}$. The properties of the iron line
complex are here given for the Fe{\sc i}+Fe{\sc xxv}+Fe{\sc xxvi} plus
the $\simeq$8.1~keV line
deconvolution. All lines refer to K$_{\alpha}$ transitions, unless
otherwise specified.}
\label{tab3}
\vspace{0.05in}
\begin{center}
\begin{tabular}{lccc} \hline \hline
Parameter & {\verb!br!} & {\verb!mk!} & Identification \\ \hline
${\rm E^{(1)}}$ &$0.59 \pm 0.04$ & $0.54 \pm 0.04$ & O{\sc vii} \\
${\rm EW^{(1)}_{tot}}$ & $80 \pm^{90}_{40}$ & $ 70 \pm 30$ &  ($1.0 \times 10^{-3}$) \\
${\rm EW^{(1)}_{sc}}$ &$340 \pm^{380}_{150}$ & $ 310 \pm 140$ & \\[0.20 cm]
${\rm E^{(2)}}$ &$0.78 \pm^{0.06}_{0.04}$ & 0.78$^{\dag}$ & Fe-L \\
${\rm EW^{(2)}_{tot}}$ & $100 \pm 30$ & $< 50$ & \\
${\rm EW^{(2)}_{sc}}$ & $380 \pm 110$ & $< 180$ & \\[0.20 cm]
${\rm E^{(3)}}$ & $0.95 \pm^{0.04}_{0.02}$ & $0.95 \pm 0.03$ & Ne{\sc ix} \\
${\rm EW^{(3)}_{tot}}$ & $150 \pm^{50}_{40}$ & $90 \pm 30$ & ($3.3 \times 10^{-4}$) \\
${\rm EW^{(3)}_{sc}}$ & $480 \pm^{170}_{140}$ & $320 \pm 100$ & \\[0.20 cm]
${\rm E^{(4)}}$ &$1.32 \pm^{0.06}_{0.05}$ & $1.32 \pm 0.05$ & Mg{\sc xi} \\
${\rm EW^{(4)}_{tot}}$ &$80 \pm 30$ & $70 \pm 30$ & ($8.7 \times 10^{-5}$) \\
${\rm EW^{(4)}_{sc}}$ & $180 \pm 70$ & $170 \pm 70$ & \\[0.20 cm]
${\rm E^{(5)}}$ &$1.88 \pm^{0.03}_{0.02}$ & $1.88 \pm 0.03$ & Si{\sc xiii} \\
${\rm EW^{(5)}_{tot}}$ &$180 \pm^{40}_{30}$ & $170 \pm 40$ & ($6.9 \times 10^{-5}$) \\
${\rm EW^{(5)}_{sc}}$ &$280 \pm^{70}_{40}$ & $280 \pm 60$ & \\[0.20 cm]
${\rm E^{(6)}}$ &$2.46 \pm^{0.03}_{0.04}$ & $2.45 \pm 0.04$ & S{\sc xv} \\
${\rm EW^{(6)}_{tot}}$ & $160 \pm 40$ & $160 \pm 40$ & ($3.1 \times 10^{-5}$) \\
${\rm EW^{(6)}_{sc}}$ &$220 \pm 50$ & $210 \pm 50$ & \\[0.20 cm]
${\rm E^{(7)}}$ & \multicolumn{2}{c}{$6.7^{\dag}$} & Fe{\sc xxv} \\
${\rm EW^{(7)}_{tot}}$ & \multicolumn{2}{c}{$1000 \pm 200$} & ($5.4 \times 10^{-5}$) \\
${\rm EW^{(7)}_{sc}}$ & \multicolumn{2}{c}{$3000 \pm 600$} & \\[0.20 cm]
${\rm E^{(8)}}$ & \multicolumn{2}{c}{$6.96^{\dag}$} & Fe{\sc xxvi} \\
${\rm EW^{(8)}_{tot}}$ & \multicolumn{2}{c}{$530 \pm^{140}_{170}$} & ($2.5 \times 10^{-5}$) \\
${\rm EW^{(8)}_{sc}}$ & \multicolumn{2}{c}{$1500 \pm^{400}_{500}$} & \\[0.20 
cm]
${\rm E^{(9)}}$ & \multicolumn{2}{c}{$8.1 \pm 0.2$} & Fe{\sc xxvi} 
(K$_{\beta}$) \\
${\rm EW^{(9)}_{tot}}$ & \multicolumn{2}{c}{$230 \pm 130$} & ($7.0 \times 10^{-6}$ \\
${\rm EW^{(9)}_{sc}}$ & \multicolumn{2}{c}{$600 \pm 300$} & \\[0.20 cm]
${\rm E^{(10)}}$ & \multicolumn{2}{c}{$6.4^{\dag}$} & Fe{\sc i}--{\sc xvi} \\
${\rm EW^{(10)}_{tot}}$ & \multicolumn{2}{c}{$1000 \pm 140$} & ($5.5 \times 10^{-5}$) \\
${\rm EW^{(10)}_{cr}}$ & \multicolumn{2}{c}{$1600 \pm 200$} & \\
\hline \hline
\end{tabular}
\end{center}
\noindent
\\
\hspace{1.5 cm}$^{\dag}$fixed
\end{footnotesize}
\end{table}

\begin{figure*}
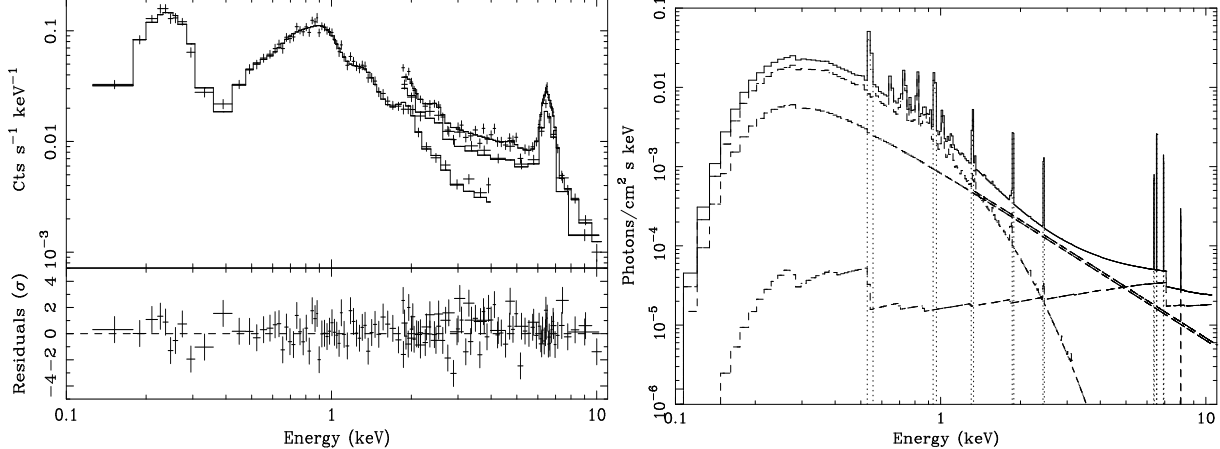

\begin{center}
\epsfig{figure=fig4.ps,height=8.0cm,width=6.0cm,angle=-90}
\epsfig{figure=fig5.ps,height=8.0cm,width=6.0cm,angle=-90}
\caption{{\it Left panel}: Spectra and residuals
in units of standard deviations when
the best-fit model (double reflector model plus optically thin plasma)
is applied to the broadband NGC~1068 spectrum (only LECS and MECS data are
shown for clarity). Each data point has a signal-to-noise ratio $>10$.
{\it Right panel}: inferred best fit model. The {\it Solid line} represents
the total spectrum, the {\it dashed lines} the continuum spectral
components and the {\it dotted lines} the emission lines}
\label{fig9}
\end{center}
\end{figure*}

When the soft excess is modeled with an optically
thin plasma, the $\chi^2$ is rather good (${\rm \chi^2_{\nu} = 1.02}$).
The addition of a further thermal component with a different temperature
is not statistically
required by the data (${\rm \Delta \chi^2 < 0.1}$).
Note that the nuclear
power-law index  remains consistent with that typically observed
in Seyfert~1 galaxies, although formally slightly steeper than the value
obtained when the high-energy spectrum alone is analyzed.
The thermal component best-fit abundance is strongly sub-solar
(${\rm \simeq 3\%}$), the temperature is $\simeq$440~eV and
the unabsorbed total luminosity is $\sim$9.3$\times 10^{41}$~erg~s$^{-1}$
(0.1--10~keV).
The best-fit line  centroid energies are 0.54, 0.95, 1.32, 1.88, 2.45~keV,
and are consistent (within the statistical uncertainties)
with K$_{\alpha}$ fluorescence of
He-like oxygen, neon, magnesium, silicon and sulphur
({\it i.e.}: O{\sc vii}, Ne{\sc ix},
Mg{\sc xi}, Si{\sc xii}, S{\sc xv}) (see Table~\ref{tab3}). 
The best-fit results and model are shown in
Figure~\ref{fig9}. The thermal component contributes substantially only
to the iron-L complex. The soft X-ray lines
EW against the {\it total} continuum are in the range 70--180~eV,
while against the {\it scattered} continuum are in the range
130-340~eV.

The total observed fluxes in the 0.1--2 and 2--10~keV bands
are 1.11 and ${\rm 0.52 \times 10^{-11}}$~erg~cm$^{-2}$~s$^{-1}$
in the 0.1--2 and 2--10~keV bands,
respectively. They correspond to
unabsorbed rest frame luminosities
of 1.31 and ${\rm 0.38 \times 10^{42}}$~erg~s$^{-1}$.
The relative fluxes (2--10~keV band) of the Compton reflection and
scattering components are 1.9 and $2.2 \times 10^{-12}$~erg~s$^{-1}$~cm$^{-2}$,
corresponding to unabsorbed luminosities of 1.34 and
$1.55 \times 10^{41}$~erg~s$^{-1}$, respectively.

The abundance of the best-fit {\verb!mekal!} model 
in the above scenario is very low and
the contribution to the observed soft X-ray emission lines
is therefore
in most cases negligible. We, therefore,  repeated the fit, assuming
this time that the soft excess is simply due to a featureless 
continuum component, either a thermal bremsstrahlung or a power law
(even though these models could not be physically plausible).
The best-fit
results and parameters are shown again in Tables~\ref{tab1} and~\ref{tab3}.
In the thermal bremsstrahlung case, the main difference is that a
line at ${\rm E \simeq 0.78}$~keV is significantly required, to account
for the missing {\verb!mekal!} contribution to the iron-L complex
(see Figure~\ref{fig9}). This new line has an EW of $\simeq$100~eV
(380~eV) against the total (scattered) continuum. Not surprisingly,
the other best fit parameters are very close to those obtained in the optically
thin plasma scenario. By contrast, a much worse fit is obtained if instead
a power-law 
is used to model the soft excess (only an upper limit
on the scattering fraction can be set in this case). The fit is unacceptable at
$\simgt 98\%$ level of confidence (${\rm \chi^2_{\nu} = 1.16}$
for 442 dof) and the steeper index obtained for the intrinsic nuclear emission
(${\rm \Gamma \simeq 2.3}$) 
produces a systematic underestimate of the PDS counts.

It is worth noticing that an unacceptable fit is obtained (${\rm
\chi^2 = 900.1/448}$~dof) if we abandon the double-reflector scenario
and assume that the soft excess (above
a bare Compton reflection) {\it and}
the  lines from ionized species are accounted for by several
single-temperatures optically thin regions.
Similar results are obtained if one adopts a single
multi-temperature emission plasma. This provides a further indirect
confirmation of the validity of the high-energy spectral deconvolution.

\section{Circinus Galaxy}
 
\subsection{BeppoSAX data}

The ASCA 0.5--10~keV spectrum of the Circinus Galaxy is well fitted
by a bare Compton-reflection component, a soft excess and a
system of emission lines (Matt et al. 1996b). The BeppoSAX observation
has permitted to discover that
at energies above $\sim$10~keV the
primary continuum emerges, suggesting that the nuclear region is seen
through a screen of absorbing matter with
${\rm N_H \sim 4 \times 10^{24}}$~cm$^{-2}$.
The reader is referred to M99 for a detailed study of the high-energy
spectrum. In the following we will
focus on the soft excess and emission line system.
Only LECS and MECS data will be considered in the following, 
since PDS data are largely affected by the transmitted
component, whose contribution is instead negligible in the 0.1--10~keV band.
For the data reduction see M99. 
Source spectra have been extracted from circular regions of 2' radius
around the source image centroid, to avoid contamination from a
serendipitous source in the field of view (M99).

The available
statistics below 2~keV is limited by the relatively large Galactic
absorption (${\rm N_{H,Gal} \sim 3 \times 10^{21}}$~cm$^{-2}$, Dickey \&
Lockman 1990). There is therefore less room than in NGC~1068
for deconvolving the soft X-ray spectral complexity
and/or to perform a detailed line spectroscopy 

\begin{figure}
\begin{center}
\epsfig{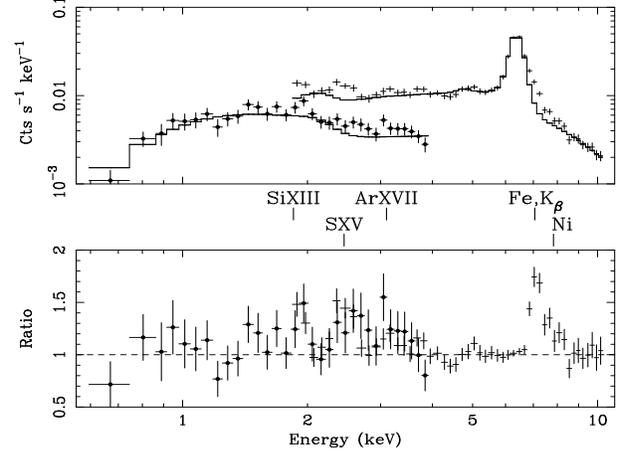}
\caption{Spectra ({\it upper panel}) and residuals in units of
standard deviations ({\it lower panel}), if the double-reflector
continuum~+~one Gaussian iron fluorescent K$_{\alpha}$ line model is
applied to the LECS ({\it filled circles}) and MECS
({\it crosses}) spectra of Circinus Galaxy. Most
likely identification of the residual line features are labeled.
All lines refers to K$_{\alpha}$ transitions if not otherwise stated}
\label{fig10}
\end{center}
\end{figure}

A rather bad fit is obtained if a double-reflector
continuum plus an iron 
K$_{\alpha}$ fluorescent line is used
(${\rm \chi^2 = 594.2/78}$~dof, see Figure~\ref{fig10}).
Most of the residuals are due to several unaccounted emission lines.
We then added narrow lines one-by-one, until required in terms
of F-test statistics at 99\% confidence level. Eventually, six lines
were needed, with best-fit centroid energies of 1.87, 2.44, 3.14, 6.45, 
7.09,
and 7.9~keV (see Table~\ref{tab6}).
\begin{table}
\begin{footnotesize}
\centering
\caption{Same as Table~3 for Circinus Galaxy.}
\label{tab6}
\vspace{0.05in}
\begin{center}
\begin{tabular}{lcc} \hline \hline
${\rm E^{(1)}}$ &$1.87 \pm^{0.04}_{0.05}$  & Si{\sc xiii} \\
${\rm EW^{(1)}_{tot}}$ & $120 \pm^{60}_{50}$  & ($3.4 \times 10^{-5}$) \\
${\rm EW^{(1)}_{sc}}$ &$160 \pm^{80}_{60}$  & \\[0.20 cm]
${\rm E^{(2)}}$ &$2.43 \pm^{0.05}_{0.04}$  & S{\sc xv} \\
${\rm EW^{(2)}_{tot}}$ & $190 \pm 50$  & ($4.0 \times 10^{-5}$) \\
${\rm EW^{(2)}_{sc}}$ & $260 \pm 70$  & \\[0.20 cm]
${\rm E^{(3)}}$ & $3.14 \pm^{0.11}_{0.09}$  & Ar{\sc xvii} \\
${\rm EW^{(3)}_{tot}}$ & $90 \pm^{40}_{50}$  & ($1.5 \times 10^{-5}$) \\
${\rm EW^{(3)}_{sc}}$ & $140 \pm^{60}_{80}$  & \\[0.20 cm]
${\rm E^{(4)}}$ &$6.446 \pm^{0.012}_{0.013}$  & Fe{\sc xvii}--{\sc xix} \\
${\rm EW^{(4)}_{tot}}$ &$2250 \pm 100$ & ($3.19 \times 10^{-4}$) \\
${\rm EW^{(4)}_{cr}}$ & $2850 \pm 130$ & \\[0.20 cm]
${\rm E^{(5)}}$ &$7.08 \pm 0.06$ & ``Neutral'' iron (K$_{\beta}$) \\
${\rm EW^{(5)}_{tot}}$ &$500 \pm^{80}_{90}$ & ($6.1 \times 10^{-5}$) \\
${\rm EW^{(5)}_{cr}}$ &$640 \pm^{100}_{110}$  & \\[0.20 cm]
${\rm E^{(6)}}$ &$7.9 \pm^{0.2}_{0.3}$ & Fe{\sc xxv} (or Ni K$_{\alpha}$) \\
${\rm EW^{(6)}_{tot}}$ & $180 \pm 90$  & ($1.6 \times 10^{-5}$) \\
${\rm EW^{(6)}_{sc}}$ &$700 \pm 300$  & \\
${\rm EW^{(6)}_{cr}}$ &$240 \pm 120$  & \\ \hline \hline
\end{tabular}
\end{center}
\end{footnotesize}
\end{table}
The most likely identifications for the first five lines
are K$_{\alpha}$ fluorescence
of Si{\sc xiii}, S{\sc xv}, Ar{\sc xvii}, Fe{\sc xvii}--{\sc xix}, and
Fe$<${\sc xvii} K$_{\beta}$, respectively. 
Most of them had already been measured in the ASCA observation
of Circinus Galaxy, with comparable EW (Matt et al. 1996b).
It is worth noticing that
the $\simeq$7.09~keV line is inconsistent, given the
uncertainties, with being produced by fluorescence of Fe{\sc xxvi},
resolving an ambiguity that was still present in the ASCA data.
The huge (${\rm EW \simeq 2.3}$~keV) line from neutral or mildly ionized iron
is well consistent with being produced in the same cold reflector, whose
emission dominates the 2--10~keV spectrum (Matt et al. 1996b).

The identification of the $\simeq$7.9~keV line is not straightforward.
In principle, it could be associated with K$_{\beta}$ fluorescence
from He-like iron. If this is the case, one should expect
a substantial contribution from the K$_{\alpha}$ line of this
ion as well. Only a 90\% upper limit of 52~eV
can be set on the EW of a 6.7~keV line (against the total continuum),
implying a K$_{\beta}$/K$_{\alpha}$ intensity ratio $>$0.22. Alternatively,
the line can be due to  K$_{\alpha}$ fluorescence of nickel.
The implied ionization state is still rather high ($>$Ni{\sc xx}).
Again, one should expect for consistency
a significant line contribution
from K$_{\alpha}$ fluorescence of H-like iron at 6.96~keV.
The 6.45~keV K$_{\alpha}$
line profile is narrow, the 90\% upper
limit on its intrinsic width being only 60~eV.
This excludes in principle any significant blending with lines produced by
ions more ionized than Fe{\sc xx}.
The 90\% upper limit on the EW (against the scattering continuum only) of a
6.96~keV narrow line is 530~eV, implying an implausible value
of the ionized nickel versus ionized iron line ratio  $\simgt$0.65.
The accuracy
in the MECS instrumental gain reconstruction is $\simlt 0.3 \%$, and
therefore comparable with or
even slightly worst than the statistical uncertainties, but not enough to
confuse the observed K$_{\alpha}$ iron line with that expected from an
H-like stage. Future high resolution, large sensitivity observations 
are needed to
allow a more definite conclusion on the identification
of such a feature. 

The ratio between
the neutral/mildly ionized iron
K$_{\beta}$ and K$_{\alpha}$ intensities is $0.19 \pm 0.04$, therefore
slightly higher than expected from neutral matter (0.11, Kikoin 1976).
The K$_{\beta}$ line energy overlaps with the strong absorption
edge associated with the Compton reflection component. Any small
change in the detailed shape of the edge (due, for example, to a low
degree of ionization of the reflector that might smear the edge
shape) could have a significant effect on the K$_{\beta}$ line
intensity. We have tested this possibility using the
model {\verb!pexriv!} in {\sc Xspec} (Magdziarz \& Zdziarski 1995),
which takes self-consistently into account the 
ionization structure of the reflector. We have fixed the temperature
to the value $10^5$~K to avoid over-fitting the data, thus leaving
only the ionization parameter (${\rm \xi}$) free.
The improvement in the quality of the fit is
very low (${\rm \Delta \chi^2 = 0.8}$)
and the ionization parameter is very
loosely constrained [${\rm \log(\xi) = 0.0 \pm^{1.8}_{1.4}}$].
Nonetheless,
the K$_{\beta}$/K$_{\alpha}$ intensity ratio becomes $0.15 \pm^{0.08}_{0.05}$,
consistent with the expectations. This may be a further piece of evidence
in favor of the idea that the cold reflector in Circinus Galaxy
is actually mildly ionized. Alternatively, if the iron abundance is larger
than solar, the intensity of the
7.09~keV line could be overestimated to compensate for the missing
continuum photons above the photoelectric edge of neutral iron.
Interestingly enough, an iron overabundance by a factor of about three
is independently suggested by the EW of the K$_{\alpha}$ line (see Sect.~5.1).
However, if we leave the iron
abundance  free to vary in the fit, no improvement is obtained,
the K$_{\beta}$ versus
K$_{\alpha}$ intensity ratio remaining basically unchanged, and
${\rm Z_{Fe} = 0.9 \pm^{1.4}_{0.3}}$.

The $\simeq 3.14$~keV
Ar{\sc xvii} line observed by BeppoSAX escaped detection in the
ASCA data. We stress here that the exposure time of the BeppoSAX
observation is about four times higher than the ASCA one. On the other hand,
two lines seen by ASCA have not been detected by the
LECS (at $\simeq$0.8~keV and at  $\simeq 1.33$~keV (Mg{\sc x}--{\sc xi})),
not surprisingly, given the larger ASCA sensitivity at these energies.
The corresponding BeppoSAX 90\% upper limits on the equivalent widths
(121 and 53~eV,
respectively) are, however, consistent with the ASCA values (Matt et al.
1996b).

\begin{figure*}
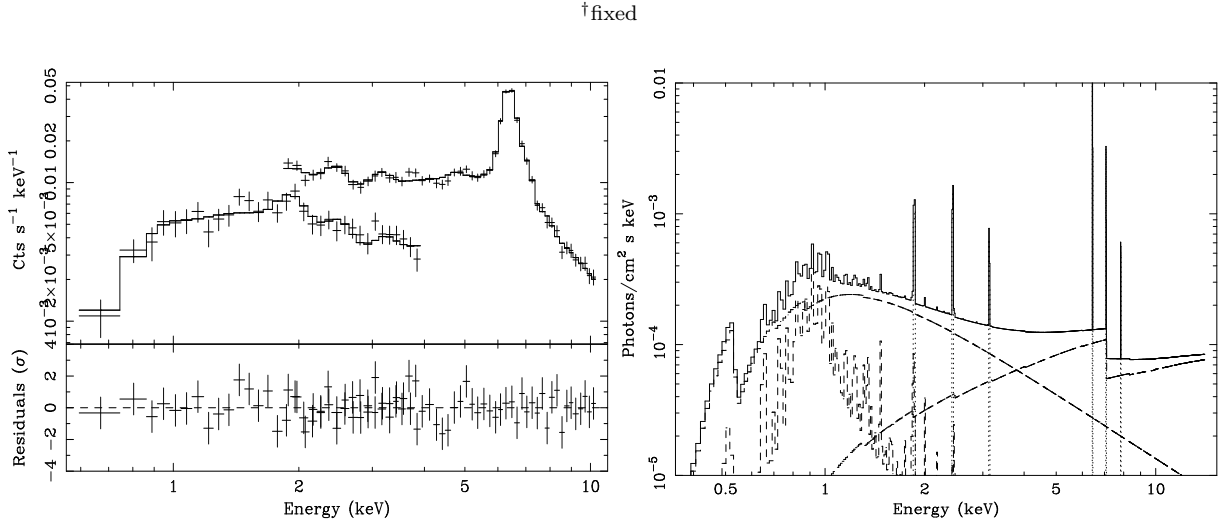

\begin{center}
\epsfig{figure=fig7.ps,height=8.0cm,width=6.0cm,angle=-90}
\epsfig{figure=fig8.ps,height=8.0cm,width=6.0cm,angle=-90}
\caption{{\it Left panel}: Spectra and residuals
in units of standard deviations when
the best-fit model (double reflector continuum plus emission lines)
is applied to the broadband Circinus Galaxy spectrum.Each data point has a 
signal-to-noise ratio $>10$.
{\it Right panel}: inferred best fit model. The {\it Solid line} represents
the total spectrum, the {\it dashed lines} the continuum spectral
components and the {\it dotted lines} the emission lines.}
\label{fig11}
\end{center}
\end{figure*}

\begin{table*}
\begin{footnotesize}
\centering
\caption{Best-fit results of the Circinus galaxy 0.1--10~keV fits
for the continuum parameters in: (first row) the double reflector 
continuum plus emission lines model and (second row) the same model plus a
thermal plasma.}
\label{tab4}
\vspace{0.05in}
\begin{center}
\begin{tabular}{lcccccc} \hline \hline
Model & ${\rm N_H}$ & ${\rm \Gamma}$ & ${\rm N_{sc}/N_{pexrav}}$ & ${\rm kT}$ 
& ${\rm Z}$ & ${\rm \chi^2/}$~dof \\mm
& ${\rm 10^{20}}$~cm$^{-2}$ & & (\%) & (keV) & (\%) & \\ \hline
{\verb!po!} & $25 \pm 7$ & $1.6 \pm 0.2$ & $3.1 \pm^{2.1}_{1.1}$ & & & 58.7/67 
\\
{\verb!po+mekal!} & $80 \pm^{60}_{50}$ & $1.6 \pm^{0.2}_{0.3}$ & $3.8 
\pm^{4.7}_{1.6}$ & $0.5 \pm^{0.7}_{0.3}$ & 100$^{\dag}$ & 54.9/65 \\ \hline 
\hline
\end{tabular}
\end{center}
\noindent
$^{\dag}$fixed
\end{footnotesize}
\end{table*}
In Figure~\ref{fig11} the spectra, best-fit model (double reflector continuum
plus emission lines) and residuals are shown in Figure~\ref{fig11}.
As already observed by M99
the best-fit photon index of the primary continuum ($\simeq1.6$;
see Table~\ref{tab4}),
is slightly flatter than that typically observed in Seyfert~1 galaxies,
albeit not inconsistent given the relatively large
uncertainties. 
The total observed fluxes in the 0.1--2 and 2--10~keV
energy bands are $8.7 \times 10^{-13}$ and
$1.37 \times 10^{-11}$~erg~cm$^{-2}$~s$^{-1}$, corresponding
to unabsorbed luminosities of $2.9 \times 10^{40}$ and
$1.34 \times 10^{41}$~erg~s$^{-1}$, respectively. The 2--10~keV band fluxes
of the Compton reflection and
scattering components are 6.3 and $3.0 \times 10^{-12}$~erg~s$^{-1}$~cm$^{-2}$,
corresponding to unabsorbed luminosities of 6.0 and
$2.9 \times 10^{40}$~erg~s$^{-1}$.

If a single temperature {\verb!mekal!} model
is added to the best--fit double reflector continuum plus lines
model, a marginal
improvement in the quality of the fit is obtained
(${\rm \Delta \chi^2 = 3.8}$ for two more free parameters).
The Compton-reflection, scattering and line properties are not substantially
affected by the inclusion of this further component. 
The unabsorbed 0.1--10~keV luminosity corresponding to the best-fit nominal
temperature ($\simeq$500~eV) is
$1.4 \times 10^{40}$~erg~s$^{-1}$, while the abundance is basically
unconstrained. 

No significant improvement in the quality of the fit is obtained if
the optically thin plasma is substituted by
a bremsstrahlung
or if the soft X-ray power-law spectral index is left free to
vary. Again, neglecting the double-reflector scenario and assuming that the
whole soft excess and the ionized lines are due to a single or
multi-temperature optically thin plasma yields an unacceptable 
${\rm \chi^2}$ (98.0/69~dof).

\subsection{HRI data}

Contrary to NGC~1068, little can be found in the
literature about the X-ray imaging of the Circinus
Galaxy. For this reason, we present here the
results of the only ROSAT/HRI observation
(spatial resolution $\simeq$10'' Full Width Half
Maximum and 2-band spectral resolution 
in the nominal 0.1--2.4~keV
energy bandpass), whose data
are publicly available.
The observation was performed
from September 14 to 15, 1995 for a total
exposure time of about 4200 seconds. Data were
retrieved as event list from the HEASARC public archive, images were
extracted with {\sc Xselect} and analyzed using
the {\sc Ximage} package (Giommi et al. 1991). 
PHA channels between 2 and 
12 (approximately 0.1 and 2~keV) were used to optimize 
the signal to noise ratio; only
data with a good attitude reconstruction were used. 

Four X-ray sources are detected at
more than $3\sigma$ level  in the HRI field of view 
\begin{figure}
\begin{center}
\epsfig{figure=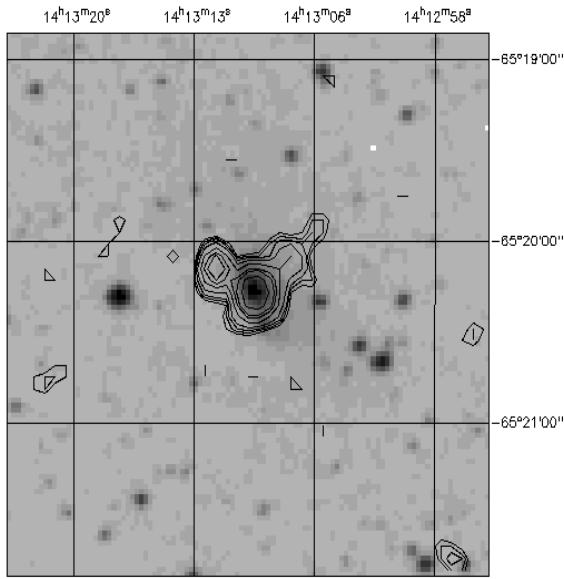,height=8.0cm,width=8.0cm,angle=-90}
\caption{Zoom of the central 10 arcmin region of ROSAT/HRI 
field of view containing the Circinus Galaxy field
({\it contours}),superimposed to an optical plate}
\label{fig13}
\end{center}
\end{figure}
(see Table~\ref{tab7}  and also Figure~\ref{fig13}).
\begin{table}
\begin{footnotesize}
\centering
\caption{Serendipitous sources detected in the ROSAT/HRI observation
of the sky region containing the Circinus Galaxy}
\label{tab7}
\vspace{0.05cm}
\begin{center}
\begin{tabular}{lccccc} \hline \hline
Source \#  & R.A.(J2000) & Dec.(J2000) & Count~Rate \\
          &            &             & (s$^{-1}$) \\ \hline
 1        &14 13 09.5 & -65 20 14.8 & $0.0128\pm0.0019$ \\
 2        &14 13 07.5 & -65 20 06.4 & $0.0030\pm0.0010$ \\
 3        &14 13 12.0 & -65 20 07.5 & $0.0039\pm0.0011$ \\
 4        &14 12 38.8 & -65 23 30.7 & $0.0167\pm0.0022$  \\ \hline \hline
\end{tabular}
\end{center}
\end{footnotesize}
\end{table}
Source \#1 is very close to the optical position of 
Circinus galaxy nucleus, allowing the identification of this X-ray source
with the AGN.
Sources \#2 and
\#3 are 30--40 arcsec away from the nuclear region, while source \#4
is located about 5 arcmin from the galaxy in the S-W direction. This last
source was also detected  by ASCA (Matt et al. 1996b) and BeppoSAX
(M99).

No extended emission is detected around the nucleus
If we subtract the two sources closer to the nucleus,
the residual nuclear emission is consistent with the PSF of the instrument
(Davis et al. 1996). We stress, however,
that the available exposure time is rather low and more statistics
is needed before reaching a firm conclusion on this point. The flux of
source \#1 is $\simeq 4.2 \times 10^{-13}$~erg~cm$^{-2}$~s$^{-2}$,
which is about half of that observed by BeppoSAX. Sources \#2 and
\#3 lie well within the extraction radius
used for the BeppoSAX observation and
could in principle contaminate the soft X-ray flux. If they are not
variable, such a contamination could be as high as 25\%, and therefore
account for a substantial fraction of the galaxy soft excess.
Although the possibility that the
soft X-ray flux in the  BeppoSAX data
is entirely to one of these serendipitous sources cannot 
be ruled out {\it a priori}, it
would however require that at
least one of these sources varies by almost an order of magnitude up to
a luminosity level $\simgt 10^{40}$~erg~s$^{-1}$, while keeping
coincidentally a $\Gamma \simeq 1.5$. Although we cannot reject this
hypothesis on the basis of BeppoSAX (or ASCA) data alone, we consider it
almost unlikely and do not discuss it in the following.

\section{Discussion}

\subsection{X-ray lines}

The BeppoSAX observations of NGC~1068 and Circinus galaxy have shown
line-rich X-ray spectra. The lines are generally associated with
He- or H-like ions of elements heavier than oxygen, the iron-L complex,
and K$_{\alpha}$ (and, possibly ${\rm K_{\beta}}$) fluorescent lines from
neutral or mildly ionized iron. Similar lines had already been observed
by BBXRT and ASCA, and most often
associated with reprocessing of the primary radiation in the nuclear
environment (Marshall et al. 1993; Turner et al. 1996b;
Netzer \& Turner 1997, M97).
The energy resolution of the imaging instruments on-board
BeppoSAX is not as good as that of the CCD on board ASCA and this limits its
capability of deconvolving 
the K-shell iron line complex. However, at least two new results
emerge from the BeppoSAX spectroscopy: i) the detection 
of emission lines around 8~keV, where BeppoSAX instruments
can take advantage of a better effective area and comparable or lower
background than the ASCA SIS; ii) the detection of a $\simeq$0.57~keV
line in NGC~1068,
which is most likely due to the K$_{\alpha}$ fluorescence from
He-like oxygen. The lack of the latter feature in previous
spectra had forced to
assume {\it ad hoc} oxygen-poor plasma (Marshall et al. 1993;
Netzer \& Turner 1997) in modeling the NGC~1068 warm mirror.
This problem has now been overcome,
given the accurate calibration of the LECS instrument on-board BeppoSAX
around 0.5~keV (cf. Haardt et al. 1998; Orr et al. 1998).

A common feature of the spectra presented here is that the line contribution
from any thermal plasma is negligible, with the possible exception
of the iron-L complex (cf. Figure~\ref{fig9} and \ref{fig11}). This
has implications on the nature of this spectral component which we
tentatively associate with the nuclear starburst and  discuss
in the next section. In this section, instead,  we will focus on 
the properties of
all the other lines, which we assume to originate as 
fluorescence/recombination or
resonant scattering from the cold and warm reflectors.

K$_{\alpha}$ (and K$_{\beta}$)
fluorescent lines from neutral or mildly ionized iron are
likely to originate in the cold reflector. In Circinus Galaxy,
the lack of ionized iron lines allows a relatively unambiguous
identification and deconvolution of the iron emission line complex. The
observed K$_{\alpha}$ EW of $\simeq$2.3~keV is in good agreement with
the one measured by ASCA (M97) and corresponds to a value of $\simeq$2.8~keV
if measured against the Compton-reflection continuum alone. This
value, coupled with the relatively low inclination angle
(${\rm \imath \simlt 40^{\circ}}$) derived
from the hard X-ray analysis (M99), implies an iron overabundance
by at least a factor of 3 (MBF96, Matt et al. 1997b).
A K$_{\beta}$ line is detected as well, with a ratio to the K$_{\alpha}$ 
line consistent with that expected
on theoretical grounds, especially when a possible mild
ionization of the reflecting medium is taken into account. The
situation of NGC~1068 is far more complex. The limited MECS energy resolution
does not allow to unambiguously deconvolve the line
complex. We have assumed in the above analysis the reasonable hypothesis
that the line is given by the superposition of a ``neutral'', a He-like
and a H-like component. In this hypothesis, the EW of the neutral line
against the Compton-reflection continuum is $\simeq 1.6$~keV. Following
MBF96, this 
would suggest a high inclination (in agreement with the idea that the line
comes from the same region as the water maser at
${\rm \imath \simgt 82^{\circ}}$),
but the large uncertainties on this value prevent any 
conclusion on this point.

The other lines observed correspond to He-like K$_{\alpha}$ transitions
from O, Ne, Mg, Si, S, Fe and H-like Fe. Recent works have focused on 
producing self-consistent models for the lines expected from
a plasma, photoionized by a strongly absorbed Seyfert-like continuum. In the
following we will mainly refer
to the works by Netzer \& Turner (1997) and NTG98 and define the  ionization
parameter ${\rm U_X}$ as the dimensionless
ratio between the ionizing photon flux (integrated in the 0.1--10~keV energy
band) and the electron
density (George et al. 1998). The calculations used in these papers take into
account both the fluorescence/recombination and the resonance scattering lines,
the latter process being dominant at densities $\simlt 10^{22}$~cm$^{-2}$.
We first consider the more complex case of NGC~1068.
The wide range of ionized species observed can be hardly reconciled with a
single-zone, homogeneous plasma. Assuming a density
${\rm N_{H,warm} = 2.5 \times 10^{22}}$~cm$^{-2}$, all the lines from 
intermediate
elements  require ${\rm 0.5 \le U_X \le 5}$, except for the oxygen one
(cf. Fig.~2 of NTG98).
On the other hand, the highly ionized iron lines require ${\rm U_X \ge 5}$.
At this high level of ionization, oxygen is likely to be at least 
in H-like state,
if not fully stripped. An intense O{\sc vii} line is 
instead still consistent with neutral iron, requiring ${\rm U_X \le 0.5}$. 
We therefore conclude that the warm mirror must be complex and structured.
At least three
components are needed in
NGC~1068, unless the oxygen line originates in the same cold reflector
as the neutral iron lines. 
In Circinus Galaxy, the lack of detection of oxygen and highly ionized iron
lines overcomes these difficulties, and the observed set of ionized lines
can be easily explained with a single-zone scatterer with intermediate
ionization parameter. However, while any contribution from a strongly
ionized scatterer is ruled out,
oxygen (and magnesium) lines are likely to have missed detection
due to the higher Galactic
neutral absorption along the line of sight. Actually,
ASCA revealed a Mg{\sc ix} line, with an EW of about 90~eV (Matt et al.
1996b), which is consistent with the upper limit derived from the BeppoSAX
data.

\subsection{Some numbers on the reflectors}

\begin{figure}
\begin{center}
\epsfig{figure=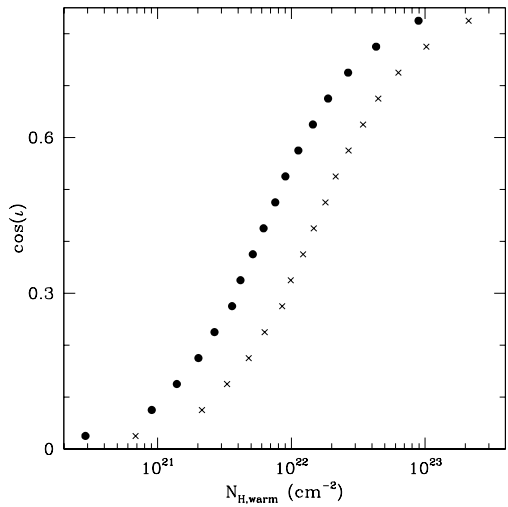,height=7.0cm,width=7.0cm}
\caption{${\rm \cos \imath}$ vs. ${\rm N_{H,warm}}$ relation in
NGC~1068 ({\it crosses}) and Circinus Galaxy ({\it filled circles})}
\label{fig12}
\end{center}
\end{figure}

For a geometry such as  that adopted
by Ghisellini et al. (1994), the amount of cold reflection depends mainly on 
the
inclination angle, while that of the ionized scattering depends almost
entirely on the optical depth (or column density) of the scattering 
material (MBF96). From the 2--10~keV flux
ratio of the warm vs cold reflection, we may therefore derive a relation
between ${\rm \imath}$ and ${\rm N_{H, warm}}$, which is shown in
Fig.~\ref{fig12}.

In the NGC~1068 case, the source is likely to be observed at very high
inclination angles, as also implied by water maser measurements (Greenhill et al.
1996)
and by the amount of cold reflection continuum measured (M97).
Then ${\rm N_{H, warm}}$ is constrained to be lower than, say, a 
few$\times$10$^{21}$ cm$^{-2}$.
An independent estimate of ${\rm N_{H, warm}}$
can be derived from the equivalent widths of the ionized lines (with respect
to the warm scattering continuum). These values ($\sim$3 keV and $\sim$1.5 keV
for the He- and H- like ions, respectively) may be compared with Figs.~4, 5
and 6 of MBF96, where the EW of these two ions are shown as a function
of ${\rm N_{H, warm}}$ (the total line EW decreases with ${\rm N_{H, warm}}$
because
the resonant scattering line becomes more and more optically thick). The values
in those figures are for solar abundances and fraction of ions equal to 1. 
The value of the EW of the cold iron line (with respect to the Compton
reflection)
implies that the iron abundance is roughly solar (MBF96). Assuming 
typical ion fractions of $\simlt$0.5, the comparison of observed and expected EW
yields ${\rm N_{H, warm}}$ values lower than 
10$^{21}$ cm$^{-2}$, even allowing for 
turbulent motions in the matter (which reduce the line opacity).
The EW of the other ionized
lines are much lower than that of iron, but they are probably calculated 
against the {\it wrong} continuum and so not much information can be gathered
from them. In any case, it 
must be stressed that the above calculation is not self consistent due to
the presence of mildly ionized material responsible
for the intermediate Z lines; instruments with much larger sensitivity and
far better energy resolution are necessary to make any further progress
in understanding the geometry and physics of the nuclear environment of
NGC~1068. 

The same calculation can be applied to Circinus Galaxy. If we
assume that the inclination angle is $\simlt 40^{\circ}$ (M99), the
density of the warm scatterer has to be higher than a few $10^{22}$~cm$^{-2}$.
This is in good agreement both with the value
derived from the S{\sc xv} line EW (NTG98)
and with the lack of detection of highly ionized iron lines (but the 90\% upper
limits on the EW of the He- or H-like iron lines are 
$\sim$1.5 keV, i.e. consistent with practically any column density).
The intrinsic luminosity is given by
$L_X (2-10 keV) \sim {L_{\rm sc} \over
\tau \Delta\Omega/4 \pi}$,
where $\Delta\Omega$ is the solid angle subtended by the visible part
of the illuminated ionized matter. From the hard X--ray spectrum, 
M99 derives $L_X (2-10 keV) \sim 10^{42}$ erg s$^{-1}$, which implies
$\tau \Delta\Omega/4 \pi \sim 0.008$.
In order to have, say, ${\rm N_{H,warm} = 3 \times 10^{22}}$~cm$^{-2}$,
$\Delta\Omega/4\pi$ should be $\sim$0.3, corresponding to
an half--opening angle of the torus of $\simeq$65$^{\circ}$.

A ${\rm N_{H,warm} \sim 10^{22}}$~cm$^{-2}$ scatterer should also imprint
absorption features in the soft X-ray spectrum. The addition of photoionization
absorption edges does not improve the quality of the fit in either NGC~1068
or Circinus~Galaxy. The upper limits
on the optical depths of O{\sc vii} and O{\sc viii} photoionization edges
are 0.35 and 0.25 in NGC~1068, respectively. The same limits are 2.7 and
1.5 in the
low Galactic latitude (and therefore highly absorbed) Circinus Galaxy.
Assuming that the scattering plasma is in the typical conditions of the
``warm absorbers'' observed in Seyfert~1 galaxies (Reynolds 1997; George et 
al. 1998), these values correspond to hydrogen  column densities
of ${\rm N_{H,warm} \simlt 5 \times 10^{21}}$~cm$^{-2}$ and
${\rm N_{H,warm} \simlt 4 \times 10^{22}}$~cm$^{-2}$, broadly in agreement
with the above diagnostics.

\subsection{On the nature of the soft X-ray continua}

In both Circinus Galaxy and NGC~1068, the soft X-rays continuum is
most convincingly explained as the superposition of two components
(even if for Circinus the evidence for a second component is marginal):
a scattered power-law, which mirrors a fraction ${\rm f_s}$
of the primary nuclear emission; and an optically thin plasma emission,
which we tentatively associate with a  nuclear
starburst. Indeed, both galaxies exhibits strong star-forming activity both in
optical and IR (Scoville 1988; Maiolino et al. 1998).

Typical temperatures are $\simeq$0.4--0.5~keV. They are slightly
lower than those (0.6--0.9~keV) observed in starburst galaxies by
ASCA (Ptak et al. 1999) and BeppoSAX (Persic et al. 1998; Cappi 1999).
David et al. (1992) studied the correlation between far-infrared (FIR)
and X-ray luminosity in starburst galaxies and derived a logarithmic
scale law. The FIR luminosity of NGC~1068 (calculated assuming the
formula (1) in David et al. 1992) is $2.2 \times 10^{44}$
(Soifer et al. 1989); it
translates into an expected 0.5--4.5~keV luminosity of
$\sim$10$^{41}$~erg~s$^{-1}$, which is roughly consistent with that 
of the thermal component {\it alone} ($3.7 \times 10^{41}$~erg~s$^{-1}$,
given also the admittedly
wide spread in the correlation. For Circinus Galaxy the integrated
IR luminosity measured ($4 \times 10^{43}$~erg~s$^{-1}$, 
Moorwood et al. 1996; Maiolino et al. 1998) corresponds
to an expected X-ray luminosity of $\sim 2.1 \times 10^{40}$~erg~s$^{-1}$,
again broadly consistent with the observations.
This evidence strengthens the
case for the identification of this soft thermal component
with the starburst. However, in Circinus~Galaxy
we cannot rule out the possibility  that the excess emission is partly
or entirely due to contaminating X-ray sources in the bulge of the galaxy.

In NGC~1068 the thermal component abundance is more than one order of magnitude
lower than solar and in agreement with the value derived from
the ASCA data (Netzer \& Turner 1997).
Similar values have been measured in starburst
galaxies observed by ASCA (Ptak et al. 1998) and attributed to the
relative weakness of the iron L complex lines. When
the X-ray spectra of these galaxies  are fitted with thermal plasmas,
and the abundances of iron and
$\alpha$-elements are decoupled, the $\alpha$-elements
abundance recovers solar or even over-solar values (Persic et al. 1998),
whereas the iron abundance remains robustly sub-solar (Cappi et al. 1999). 
This low metallicity is possibly due to iron depletion
in warm interstellar  clouds
surrounded by a dust-rich galactic environment
(Ptak et al. 1997). In principle, dilution
by featureless continua, which may be provided by an unresolved population of
accretion-driven sources, may explain this effect
as well. These sources can be binaries in the nuclear starburst environment
or in the host galaxy, spatially confused with the
soft X-ray emission.
There is  clear evidence that the nuclear stellar population
in NGC~1068 is relatively young ($\sim$5--35~Myr; Davies
et al 1998) and, therefore, a large contribution by High Mass X--ray 
Binaries (HMXRB) is expected. HMXRB are, however,
generally flatter in the intermediate X-rays than NGC~1068 and, 
if they contribute significantly to the soft X-ray flux,
their summed emission should dominate above 1~keV; this is not
observed in the data.
Alternatively, wind accreting Low Mass X-ray Binaries (LMXRB) are 
harder and soft X-ray brighter. In ``normal'' spiral galaxies like our own and
M~31 (Trincheri \& Fabbiano 1991), they indeed dominate the total X-ray
output, with an integrated X-ray luminosity in the range
$10^{39}$--$10^{41}$~erg~s$^{-1}$; the brightest sources generally
cluster around the galactic bulge. 
Of course, the soft X-ray spectrum might well be more complex than our
simple, statistical-limited two-components deconvolution. 
BeppoSAX and ASCA results, although obtained with detectors
having relatively modest energy resolution, are suggestive of a very complex
situation, which will be clarified by the forthcoming
major X-ray missions like {\it Chandra}, XMM and Astro-E, whose high resolution
soft X-ray detectors will surely shed a new light on this
subject.

\section*{Acknowledgments}

This paper has made use of linearized event files produced at the
BeppoSAX Science Data Center.
The following institutions are acknowledged for financial support:
European Space Agency (MG, research fellowship),
Italian Space Agency (GM, RM, AM, FF), MURST (GM),
Royal Society (ACF), PPARC (KI). We thank the referee the valuable comments
which helped us in clarifying several issues.

{}


\begin{thebibliography}{}

\bibitem[]{} Anders E., Grevesse N., 1989, Geochimica and Cosmochimica Acta, 
53, 197

\bibitem[]{} Antonucci R., Miller J.S., 1985, ApJ, 297, 621

\bibitem[]{} Antonucci R., 1993, ARAA, 31, 473

\bibitem[]{} Awaki H., Koyama K., Inoue H., Halpern J.P., 1991, PASJ, 43, 195

\bibitem[]{} Boella G., Butler R., Perola G.C., 1997a, A\&AS, 112, 299

\bibitem[]{} Boella G., et al., 1997b, A\&AS, 223, 327

\bibitem[]{} Cappi M., Bassani L., Comastri A., et al., 1998, A\&A, in press

\bibitem[]{} Cappi M., 1999, Adv. Sp. Res., submitted

\bibitem[]{} Cappi M., et al., 1999, MmSAIt, in press (astroph/9809325)

\bibitem[]{} Claussen M.J., Heiligman G.M., Lo K.Y., 1984, Nature, 310, 298

\bibitem[]{} Cusumano G., Mineo T., Guainazzi M., et al., 1998, A\&A, submitted

\bibitem[]{} David L.P., Jones C., Forman W., 1992, ApJ, 388, 82

\bibitem[]{} David L.P., Harnden F.R. Jr., Kearns K.E., Zombeck M.V., 1996, "The ROSAT High Resolution Imager (HRI) Calibration Report", U.S.ROSAT SDC/SAO

\bibitem[]{} Davies R.I., Sugai H., Ward M.J., 1998, MNRAS, 300, 388

\bibitem[]{} Dickey J.M., Lockman F.J., 1990, ARA\&A, 28, 215 

\bibitem[]{} Fiore F., Guainazzi M., Grandi P., 1998,
``The cookbook for BeppoSAX NFI Spectral Analysis'', in press

\bibitem[]{} Frontera F., Costa E., Dal Fiume F., Feroci M., Nicastro L., 
Orlandini M., Palazzi E., Zavattini G., A\&AS, 112, 357

\bibitem[]{} George I.M., Fabian A.C., 1991, MNRAS, 249, 352

\bibitem[]{} George I.M., Turner T.J., Netzer H., Nandra K., Mushotzky R.F., 
Yaqoob T., 1998, ApJS, 114, 73

\bibitem[]{} Ghisellini G., Haardt F., Matt G., 1994, MNRAS, 267, 743

\bibitem[]{} Giommi P., Angelini L., Jacobs P., Tagliaferri G., in "Astronomical Data Analysis Software System I"; Worral D.M., Biemesderfer J., Barnes J. eds, 1991, A.S.P.Conf.Ser., 25, 100

\bibitem[]{} Grandi P., et al., 1997, A\&A, 325, L17

\bibitem[]{} Greenhill L.J., Gwinn C.R., Antonucci R., Barvanis R., 1996, ApJ, 
472, L21

\bibitem[]{} Greenhill L.J., Ellingsen S.P., Norris R.P., Gough R.G., Sinclair 
M.W., Moran J.M., Mushotzky R.F., 1997, ApJ, 474, L103

\bibitem[]{} Haardt F., et al, 1998, A\&A, 340, 35

\bibitem[]{} Heisler C.A., Lumsen S.L., Bailey J.A., 1997, Nature, 385, 700

\bibitem[]{} Kikoin I.K. (ed), 1976, Tables of Physical Quantities, Atomizdat, 
Moscow

\bibitem[]{} Krolik J.H., Kallman T.R., 1987, ApJ, 320, 5

\bibitem[]{} Krolik J.H., Madau P., \.Zycki P.T., 1994, ApJ, 420, 57

\bibitem[]{} Iwasawa K., Fabian A.C., Matt G., 1997, MNRAS, 289, 443

\bibitem[]{} Iwasawa K., Comastri A., 1998, MNRAS, 297, 1219

\bibitem[]{} Magdziarz P., Zdziarski A.A., 1995, MNRAS, 273, 837

\bibitem[]{} Maiolino R., Ruiz M., Rieke G.H., Keller L.D., 1995, ApJ, 446, 561

\bibitem[]{} Maiolino R., Krabbe A., Thatte N., Genzel R., 1998a, ApJ, 493, 650

\bibitem[]{} Maiolino R., Salvati M., Bassani L., Dadina M., Della Ceca R., Matt G., Risaliti G., Zamorani G., 1998, A\&A, 338, 781

\bibitem[]{} Malaguti G., Palumbo G.C.C., Cappi M., Comastri A., Otani C., 
Matsuoka M., Guainazzi M., Bassani L., Frontera F., 1998, A\&A, 331, 519

\bibitem[]{} Manzo G., et al., 1997, A\&AS, 112, 341

\bibitem[]{} Marconi A., Moorwood A.F.M., Salvati M., Oliva E., 1994, A\&A, 291, 18

\bibitem[]{} Marshall F.E., et al., 1993, ApJ, 405, 168

\bibitem[]{} Matt G., Piro L., Antonelli L.A., Fink H.H., Meurs E.J.A., Perola G.C., 1994, A\&A, 292, L13

\bibitem[]{} Matt G., Brandt W.N,, Fabian A.C., 1996a, MNRAS, 280, 823 (MBF96)

\bibitem[]{} Matt G., Fiore F., Perola G.C., Piro L., Fink H.H., Grandi P., 
Matsuoka M., Oliva E., Salvati M., 1996b, MNRAS, 281, 69

\bibitem[]{} Matt G., et al., 1997a, A\&A, 325, L13 (M97)

\bibitem[]{} Matt G., Fabian A.C., Reynolds C.S., 1997b, MNRAS, 289, 175

\bibitem[]{} Matt G., Guainazzi M., Maiolino R., et al., 1999, A\&A, 341, L39
(M99, astroph/9811301)

\bibitem[]{} Miller J.S., Goodrich R.W., Mathews W.G., 1991, ApJ, 378, 47

\bibitem[]{} Moorwood A.F.M., Lutz D., Oliva E., Marconi A., Netzer H., Genzel 
R., Sturm E., De Graauw T., 1996, A\&A, 315, L109

\bibitem[]{} Mulchaey J.S., Koratkar A., Ward M.J., Wilson A.S., Whittle M., 
Antonucci R.R.J., Kinney A.L., Hurt T., 1994, ApJ, 436, 586

\bibitem[]{} Nandra K., George I.M., Mushotzky R.F., Turner T.J., Yaqoob Y., 
1997, ApJ, 477, 602

\bibitem[]{} Nandra K., Pounds K.A., 1994, MNRAS, 268, 405

\bibitem[]{} Netzer H., Turner T.J., 1997, ApJ, 488, 694

\bibitem[]{} Netzer H., Turner T.J., George I.M., 1998, ApJ, 504, 680 (NTG98)

\bibitem[]{} Nicastro F., Elvis M., Fiore F., Perola G.C., 1998, ApJ, in press
(astroph/9808316)

\bibitem[]{} Oliva E., Marconi A., Cimatti A., di Serego Alighieri S., 1998, 
A\&A 329, L21

\bibitem[]{} Orr A., Parmar A.N., Yaqoob T., Guainazzi M., 1998, Nucl. Phys. B 
(Proc. Suppl.), 69/1-3, 496

\bibitem[]{} Parmar A.N., et al., 1997, A\&AS, 122, 309

\bibitem[]{} Persic M., et al., 1998, A\&A, 339, 33

\bibitem[]{} Piro L., et al., 1998, Nucl. Phys. B (Proc. Suppl.), 69/1-3, 481

\bibitem[]{} Ptak A., Serlemitsos P., Yaqoob T., Mushotzky R., Tsuru T., 1997, AJ, 113, 1286

\bibitem[]{} Ptak A., Serlemitsos P. Yaqoob T., Mushotzky R., 1999, ApJS, in 
press (astroph/9808159)

\bibitem[]{} Reynolds C.S., 1997, MNRAS, 286, 513

\bibitem[]{} Scoville N.Z., 1988, ApJ, 327, 61

\bibitem[]{} Soifer B.T., Bohemer L., Neugebauer G., Sanders D.B., 1989, AJ, 
98, 766

\bibitem[]{} Smith D.A., Done C., 1996, MNRAS, 180, 355

\bibitem[]{} Tran H.D., 1995, ApJ, 440, 565

\bibitem[]{} Trincheri G. \& Fabbiano G., 1991, ApJ, 382, 82

\bibitem[]{} Tully R.B., 1988, ``Nearby Galaxies catalog'', Cambridge 
University Press

\bibitem[]{} Turner T.J., George I.M., Nandra K., Mushotzky R.F., 1997a, ApJ, 
488, 164

\bibitem[]{} Turner T.J., George I.M., Nandra K., Mushotzky R.F., 1997b, ApJS, 
113, 23

\bibitem[]{} Ueno S., Mushotzky R.F., Koyama K., Iwasawa K., Awaki H., Hayashi 
I., 1994, PASJ, 46, L71

\bibitem[]{} Ueno S., 1997, Ph.D. thesis, Kyoto University

\bibitem[]{} Wilson A.S., Elvis M., Lawrence A., Bland-Hawthorn J., 1992, ApJ, 
391, L75

\bibitem{} Zdziarski A.A., Johnson W.N., Done C., Smith D., McNaron-Brown K., 
1995, ApJL, 438, 63

\end{thebibliography}
\end{document}